\documentclass[12pt]{article}
\usepackage{graphicx} 
\usepackage[margin=1in]{geometry}
\usepackage{setspace}
\usepackage{graphicx}
\usepackage{cite}
\usepackage{hyperref}
\begin{document}
	\doublespacing
	\begin{center}
		{\bf \large Determining $G$ with Laser Spectroscopy to 38 ppb}\\
		Noah Bray-Ali\\
		{\em Science Synergy, Los Angeles, California, USA, 90045}\\
		2026 US NIST Precision Measurement Program Grant Proposal\\
		
		October 29, 2026\\
		
	\end{center}
	
	\bigskip
	\begin{center}
		ABSTRACT
	\end{center}
	\singlespacing
	\noindent
	A precision measurement is proposed to determine, in a couple hours of integration time, the axion Compton frequency using a modest power (3 mW) tunable external-cavity diode laser at 2458 nm as input to drive a free-space table-top Mach–Zehnder interferometer whose sensing arm passes the expanded beam-waist ($3~{\rm mm}$) light beam through a $1~{\rm T}$ strong, $40~{\rm cm}$ long dipole magnetic field created by a custom-built permanent-magnet assembly with a large but achievable ($6~{\rm mm}$) gap between poles. As the laser frequency is slowly modulated at 1 kHz through a 600 MHz wide window that is well within the 30 GHz fine-tuning range of the laser, a small but readily observable modulation appears in the dark-port optical power of the dark-fringe phase-locked interferometer due to photons converting into axions within the light beam as it passes through the magnetic field. Measuring the axion Compton frequency, $\nu_A\approx{\rm 122~THz}$, where the dark-port power modulation peaks, to within the line-width of the laser, $\Delta \nu_A=1~{\rm MHz}$, then determines $G$ to 38 ppb, a roughly 600-fold improvement, through a relation between $\nu_A$ and $G$, involving $h$, $c$, and nucleon masses. 

	
	\bigskip
	\doublespacing
	{\em Introduction.\textemdash}
	The 2019 redefinition of the SI established exact values for several fundamental constants, including the Planck constant, $h$, and the elementary charge, $e$, enabling direct realizations of electrical units through quantum effects \cite{new-si-2019,new-si-bipm,new-si,codata2022}. Theoretically established relations, such as the expression for the von Klitzing constant, $R_K=h/e^2$, then link directly observable quantities, in this case the quantized Hall resistance of a two-dimensional electron system, to fundamental physical constants \cite{laughlin1981}. The present proposed precision measurement follows a similar strategy: optical measurement of the axion \cite{weinberg,wilczek} Compton frequency, $\nu_A$, provides access to the universal gravitational constant, $G$, through a simple relation with exact and well-known fundamental physical constants: $h$, $c$, and the nucleon masses. 
	.
	
	The metrological use of the interaction between atoms and photons, as opposed to the motion of rigid bodies, has long promised to transform the precision measurement of $G$ \cite{atomic-G}. Indeed, the best present determinations of $\alpha$ \cite{alpha-berkeley,alpha-paris} derive from cold-atom matter-wave interferometers measuring the atom Compton frequency, $\nu_{\rm At}=m_{\rm At}c^2/h$, with the absorption of a single photon by an atom with mass, $m_{\rm At}$. Still, despite more than a decade of effort, stimulated in part by a 2016 US NSF Ideas Lab \cite{nsf-idea-lab}, determinations of $G$ remain artifactual \cite{G-review}, relying on tungsten source masses \cite{jila-G} and silica-fiber torsional oscillators \cite{hust-G}.
	
	
	In metrological discussions, such as the current road-map toward the redefinition of the second using optical frequency standards \cite{new-second}, the interaction between {\em axions}  and photons, on the other hand, is more often invoked as a motivation for the implementation of improvements in standards \cite{safronova2019, safronova2023}. Proposals such as the present one, which instead use axion-photon interactions to actually determine the value of a fundamental physical coupling constant, in this case $G$, have not yet entered such discussions in a significant way. The present proposal paves the way towards such axion metrology by determining the axion Compton frequency, $\nu_A$, in a laser spectroscopy precision measurement.

	\begin{table}
		\begin{center}
			\begin{tabular}{|c|c|c|l|}
				\hline
				FY&$\Delta\nu_A~(\rm MHz)$&$\Delta G/G$ (ppb)&Comment\\
				\hline
				2027&$1$&$38$&New $G$: $600\times$ more precise\\
				\hline
			\end{tabular}
			\label{table:road-map}
			\caption{{\bf Determining $G$ with Laser Spectroscopy to 38 ppb.\textemdash} Year one (FY 2027) of the proposed precision measurement uses laser spectroscopy to determine the axion Compton frequency, $\nu_A\approx122~{\rm THz}$, whose value is predicted to within 600 MHz using the current value of $G$, as shown in Eq.~(\ref{eq:mA-Tcheck}). Using the relation shown in Eq.~(\ref{eq:G-mA}), this direct experimental measurement of $\nu_A$ to 8 parts per billion (ppb), set by the laser line-width $\Delta\nu_A=1~{\rm MHz}$, then determines $G$ to 38 ppb, as shown in Eq.~(\ref{eq:Delta-G}). This represents a roughly 600-fold decrease in relative uncertainty for $G$ compared to CODATA (2022) \cite{codata2022}. }
		\end{center}
		
	\end{table}

	{\em Linking $G$ and $\nu_A$.\textemdash}The present proposal to determine $G$ to 38 ppb, a roughly 600-fold improvement over the present 22 pm relative uncertainty \cite{codata2022}, in year-one of the precision measurement, shown in Table 1, derives from the simple relationship with the axion Compton frequency, $\nu_A$, from quantum chromodynamics (QCD) \cite{brayali2021}:
	\begin{equation}
		\label{eq:G-mA}
		G=\hbar c^5x_k^{3}y_T^{10}\frac{(h\nu_A)^5}{(k\check{T})^7},
	\end{equation}
	where $\check{T}$ is the temperature of the Big Bang, $x_k=2.821~439~372~122\ldots$ is the Wien displacement constant for the peak energy in the black-body distribution \cite{codata2022}, and the dimensionless constant, $y_T=(5/84)^{1/4}$, gives the Bang temperature, $k\check{T}=(\chi_{\rm QCD})^{1/4}y_T$, in terms of the topological susceptibility, $\chi_{\rm QCD}$, of the QCD vacuum \cite{villadoro}. The key to the proposal, however, is an implicit relation that determines the Big Bang temperature, $\check{T}$, in terms of the axion Compton frequency, $\nu_A$ \cite{brayali2021}: 
	\begin{equation}
		\label{eq:baryon-frac}
		\frac{12}{\pi^2}(x_ky_T)^2=\frac{m_pc^2}{h\nu_A}e^{-x_p}J(x_p)+\frac{m_nc^2}{h\nu_A}e^{-x_n}J(x_n),    
	\end{equation}
	where the mass of the nucleon, $m_N$, with $N=p$ for proton and $N=n$ for neutron, determines the ratio, $x_N=m_Nc^2/(k\check{T})$, which in turn fixes the value of the integral, $J(x_N)=1/(2\zeta(3))\times\int_0^{\infty}dx~e^{-x}\sqrt{x(x+2x_N)}$, with $\zeta(3)=1.202~056~903~159\ldots$ \cite{zeta-three}. Inverting the simple relationship, in Eq.~(\ref{eq:G-mA}), and using the best present value, $G=6.674~30~(15)\times10^{-11}~{\rm kg^{-1}~m^{3}~sec^{-2}}$ \cite{codata2022}, with 22 ppm relative uncertainty, the implicit relation, in Eq.~(\ref{eq:baryon-frac}), then determines $\nu_A$ to 4.8 ppm and $\check{T}$ to 0.23 ppm \cite{brayali2021}:
	\begin{eqnarray}
		\label{eq:mA-Tcheck}
		\nu_A=121~946.70~(59)~{\rm GHz}, ~~~~~~~~~~k\check{T} = 36.971~526~1~(84)~{\rm MeV},
	\end{eqnarray}
	where the resulting determination of the QCD vacuum topological susceptibility, $\sqrt{\chi_{\rm QCD}}=y_T^{-2}(k\check{T})^2=5~602.597~2~(25)~{\rm MeV}^2~[0.46~{\rm ppm}]$ lands close to the best present determination, $\sqrt{\chi_{\rm QCD}}({\rm CPT})=5~690~(50)~{\rm MeV^2}~[8.8~{\rm per~ mille}]$ \cite{villadoro}, from chiral perturbation theory (CPT), but is much more precise.
	
	{\em Proposed Precision Measurement Principle.\textemdash}	The principle of the proposed measurement of the axion Compton frequency, $\nu_A\approx122~{\rm THz}$, uses the interaction between axions and photons, known as axion electrodynamics \cite{wilczek1987}, that is the basis of the long-running series of axion dark matter experiments \cite{sikivie2024,rybka2024} ($\hbar=c=1$ units):
	\begin{equation}
		\label{eq:axion-photon}
		\mathcal{L}_{A\gamma}=g_{A\gamma\gamma}\phi_A{\bf E}_{\rm NIR}\cdot{\bf B}_{\rm DC},
	\end{equation}
	where the axion-photon coupling strength is $g_{A\gamma\gamma}=0.68~(2)\times10^{-10}~{\rm GeV}^{-1}$, $\phi_A$ is the axion amplitude in real-space, and ${\bf E}_{\rm NIR}\cdot{\bf B}_{\rm DC}$ is the dot product of the dipole magnetic field, ${\bf B}_{\rm DC}$, and the near-infrared electric field, ${\bf E}_{\rm NIR}$. However, the value of the axion Compton frequency, $\nu_A\approx122~{\rm THz}$, in Eq.~(\ref{eq:mA-Tcheck}), puts the axion Compton wavelength, $\lambda_A=c/\nu_A\approx2.46~{\rm \mu m}$, in the near-infrared part of the electromagnetic spectrum, where relatively few axion dark matter experiments have been proposed \cite{terabread,lamppost,iaxo} and none yet performed. Still, a significant number of experiments have been designed, built, and some even run, looking for the effects of axion-photon coupling on near-infrared laser light passing through a dipole magnetic field, including rotating the light polarization \cite{pvlas-one,pvlas-two}, shining the light ``through a wall'' \cite{chou,alps}, and modulating the light intensity \cite{wispfi,winter}.
	
	In contrast to axion dark matter experiments \cite{rybka2024, terabread,lamppost,iaxo}, which try to convert axions from the local dark matter halo into photons, the proposed precision measurement proceeds instead along the lines of the previous and current axion electrodynamic experiments \cite{pvlas-one, pvlas-two, chou, alps, wispfi, winter}, which use the dipole magnetic field, ${\bf B}_{\rm DC}$, to {\em make} axions from the near-infrared electric field, ${\bf E}_{\rm NIR}$, with the axion electrodynamic coupling described by Eq.~(\ref{eq:axion-photon}) (SI units here and throughout the remainder):
	\begin{equation}
		\label{eq:axion-amplitude}
		\tilde\phi_A(q_A)=\left(h\nu_Ag_{A\gamma\gamma}\right)\left(\frac{{\bf E}_{\rm NIR}\cdot{\bf B}_{\rm DC}}{\mu_0c}\right)\sqrt{\frac{V}{h\Delta\nu_A}},
	\end{equation}  
	where $\tilde{\phi}_A(q_A)$ is the resulting momentum-space amplitude of the axion with frequency-wavevector four-vector, $q_A\approx(\nu_A,0,0,0)$, the volume of the near-infrared beam with waist, $w$, within the dipole magnet of length, $\ell$, is $V=\pi w^2\ell$, and the near-infrared laser line-width is $\Delta\nu_A$. In turn, the axion amplitude, $\tilde{\phi}_A(q_A)$, given in Eq.~(\ref{eq:axion-amplitude}), induces a change in the near-infrared electric field, $\Delta {\bf E}_{\rm NIR}$, using the same axion electrodynamic coupling:
	\begin{equation}
		\label{eq:Delta-E}
		\Delta {\bf E}_{\rm NIR}=\tilde{\phi}_A(q_A)\left(h\nu_Ag_{A\gamma\gamma}\right){\bf B}_{\rm DC}c\sqrt{\frac{V}{h\Delta\nu_A}}.
	\end{equation} 
	Combining the expression for the change in the near-infrared electric field, $\Delta{\bf E}_{\rm NIR}$, in Eq.~(\ref{eq:Delta-E}), with that for the momentum-space amplitude of the axion, $\tilde{\phi}_A(q_A)$, in Eq.~(\ref{eq:axion-amplitude}), yields the modulation, $\Delta P_{\rm NIR}$, in the near-infrared laser power, $P_{\rm NIR}$, which the proposed precision measurement seeks to observe when the laser frequency is tuned to $\nu_A$:
	\begin{eqnarray}
		\label{eq:Delta-P}
		\Delta P_{\rm NIR}&=&\frac{1}{4}\left(h\nu_Ag_{A\gamma\gamma}\right)^2\left(\frac{B_{\rm DC}^2V}{\mu_0h\Delta\nu_A}\right)P_{\rm NIR}\nonumber\\
		&=&12.0~(2)~{\rm fW}\left(\frac{P_{\rm NIR}}{3~{\rm mW}}\right)\left(\frac{1~{\rm MHz}}{\Delta\nu_A}\right)\left(\frac{B_{\rm DC}}{1~{\rm T}}\right)^{2}\left(\frac{w}{3~{\rm mm}}\right)^{2}\left(\frac{\ell}{40~{\rm cm}}\right),
	\end{eqnarray} 
	where the tunable near-infrared laser power, $P_{\rm NIR}=3~{\rm mW}$, and line-width, $\Delta\nu_A=1~{\rm MHz}$, at the axion Compton wavelength $\lambda_A\approx 2458~{\rm nm}$, are the specifications for the commercially available external cavity diode laser \cite{laser} whose purchase price dominates the year-one budget of the proposed precision measurement, while the length, $\ell=40~{\rm cm}$, strength, $B_{\rm DC}=1~{\rm T}$, and gap between poles, $2w=6~{\rm mm}$, are reasonable parameters for a dipole magnetic field created by a custom-built permanent magnet assembly \cite{wispfi}.
	
	\begin{figure}
		\centering
		\includegraphics[width=1.0\textwidth]{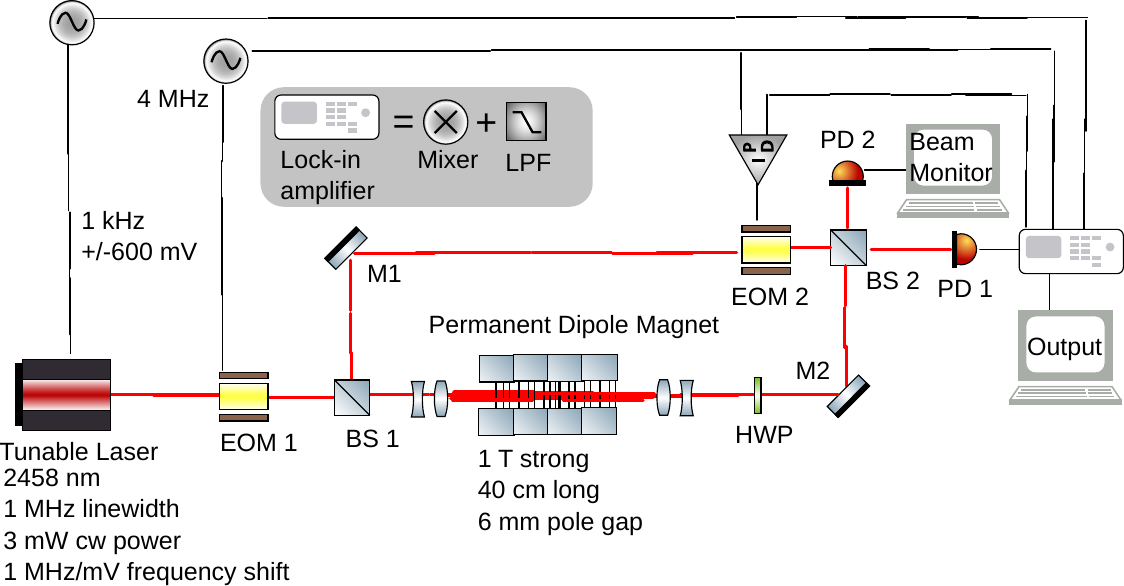}
	
		\caption{{\bf Determing $G$ with Laser Spectroscopy to 38 ppb\textemdash} Schematic of proposed Year One (FY 2027) precision measurement showing a frequency-modulated and phase-locked free-space Mach-Zehnder inteferometer for sensing modulation of the transmitted intensity of near-infrared light passing through a static magnetic field. The tunable 3 mW external cavity diode laser input at 2458 nm with 1 MHz line-width \cite{laser} is frequency-modulated at $1~{\rm kHz}$ through a $\pm600~{\rm MHz}$ window around the predicted axion Compton frequency, $\nu_A=121~946.70~(59)~{\rm GHz}$. The laser beam-waist is broadened to $w=3~{\rm mm}$ as it passes through the dipole magnetic field with strength, $B_{\rm DC}=1~{\rm T}$, and length, $\ell=40~{\rm cm}$, within a custom-built permanent magnet assembly in the ``sensing arm'' of the interferometer. A resonant electro-optical modulator (EOM 1) at 4 MHz provides the phase-lock reference and a broadband electro-optical modulator (EOM 2) in the ``control arm'' compensates phase drifts with a proportional-integral-derivative (PID) controller. Other components include mirrors (M), beam splitters (BS), a half-wave plate (HWP), photodiodes (PD), and a low-pass filter (LPF). }
	\end{figure}

	{\em Proposed Precision Measurement.\textemdash} The set-up for the proposed precision measurement, shown in Figure 1, is a phase-locked and frequency-modulated free-space Mach-Zehnder interferometer for sensing modulation of the transmitted intensity of near-infrared light passing through a static magnetic field. The interferometer takes as input a tunable external cavity diode laser beam with continuous-wave power, $P_{\rm NIR}=3~{\rm mW}$, tuned to the axion Compton wavelength, $\lambda_A\approx2458~{\rm nm}$, with line-width $\Delta\nu_A={\rm 1~MHz}$ \cite{laser}. The laser beam-waist is broadened to $w=3~{\rm mm}$ as it passes through the dipole magnetic field with strength, $B_{\rm DC}=1~{\rm T}$, and length, $\ell=40~{\rm cm}$, within a custom-built permanent magnet assembly in the ``sensing arm'' of the interferometer.
	
	The concept of the proposed precision measurement is to frequency-modulate the laser input at $1~{\rm kHz}$ through a $\pm600~{\rm MHz}$ window around the predicted axion frequency, given by Eq.~(\ref{eq:mA-Tcheck}), and then to lock-in amplify the change in the dark-fringe intensity, $\Delta P_{\rm NIR}$, in the ``dark port'' photodetector (PD 1) of the inteferometer. The phase-lock on the dark fringe is accomplished by a resonant electro-optical modulator (EOM 1) at 4 MHz that provides the phase-lock reference and by a broadband electro-optical modulator (EOM 2) in the ``control arm'' of the inteferometer that compensates phase drifts with a PID controller. In this way, the signal-to-noise ratio, ${\rm SNR}=\Delta P_{\rm NIR}\sqrt{T}/{\rm NEP}$, is determined by the integration time, $T$, and by the typical noise-equivalent power, ${\rm NEP}=200~{\rm fW}/\sqrt{\rm Hz}$, of the dark-port photodiode \cite{diode}.
	
	The proposed integration time, $T\approx2~{\rm h}$, suffices to reach the signal-to-noise ratio, ${\rm SNR}=5$, for the the axion electrodynamic modulation, $\Delta P_{\rm NIR}$, given in Eq.~(\ref{eq:Delta-P}):
	\begin{eqnarray}
		\label{eq:T}
		T&=&\left(\frac{{\rm SNR\times NEP}}{\Delta P_{\rm NIR}}\right)^2\nonumber\\
		&=&1.9~(2)~{\rm h} \left(\frac{{\rm SNR}}{5}\right)^{2}\left(\frac{\rm NEP}{200~{\rm fW}/\sqrt{\rm Hz}}\right)^{2}\nonumber\\
		&\times&\left(\frac{3~{\rm mW}}{P_{\rm NIR}}\right)^{2}\left(\frac{\Delta \nu_A}{1~{\rm MHz}}\right)^{2}
		\left(\frac{1~{\rm T}}{B_{\rm DC}}\right)^{4}\left(\frac{3~{\rm mm}}{w}\right)^{4}\left(\frac{40~{\rm cm}}{\ell}\right)^{2}.		
	\end{eqnarray}
	The resulting determination of the axion Compton frequency, $\nu_A\approx122~{\rm THz}$, then has precision, $\Delta\nu_A/\nu_A\approx8~{\rm ppb}$, set by the laser line-width, $\Delta\nu_A=1~{\rm MHz}$. In turn, the simple relationship between the universal gravitational constant, $G$, and the axion Compton frequency, $\nu_A$, shown in Eq.~(\ref{eq:G-mA}), then determines $G$ to 38 ppb, a roughly 600-fold improvement over the best present determination \cite{codata2022}:
	\begin{eqnarray}
		\label{eq:Delta-G}
		\frac{\Delta G}{G}=\left(5-7\frac{h\nu_A}{k\check{T}}\frac{\Delta k\check{T}}{\Delta \nu_A}\right)\left(\frac{\Delta \nu_A}{\nu_A}\right)\approx\left(5-\frac{7}{x_N}\right)\left(\frac{\Delta\nu_A}{\nu_A}\right)=38~{\rm ppb},
	\end{eqnarray}    
	where the value of the ratio, $x_N=m_Nc^2/(k\check{T})\approx25.4$, is set by the best present value for the Big Bang temperature, $k\check{T}\approx37.0~{\rm MeV}$, given in Eq.~(\ref{eq:mA-Tcheck}), and by the approximate logarithmic derivative, $(h\nu_A/k\check{T})(\Delta k\check{T}/\Delta \nu_A)=(1/x_N)\left(1-J'(x_N)/J(x_N)\right)\approx 1/x_N$, was evaluated using the corresponding logarithmic derivative, $J'(x_N)/J(x_N)=0.02$, of the integral, $J(x_N)$, whose expression is given below Eq.~(\ref{eq:baryon-frac}).

	{\em Conclusion.\textemdash} A feasible table-top magneto-optical precision measurement was proposed for determining the universal gravitational constant with relative uncertainty more than two orders of magnitude smaller than the best present determination in the first year of the program. The proposed experiment resembles a conventional optical spectroscopy measurement in which a tunable laser is scanned across a narrow resonance and the resulting modulation of the optical field is detected interferometrically. The directly observed resonance frequency determines the universal gravitational constant through a simple relationship involving the Planck constant, the speed of light, and nucleon masses.

\end{document}